\documentstyle[12pt]{article}
\newdimen\deltaabs \newdimen\deltaref \newdimen\deltabase
\def\kurzformat
  {\addtolength{\topmargin}{-30mm}
   \setlength{\textheight}{258mm}
   \addtolength{\oddsidemargin}{-14mm}
   \setlength{\textwidth}{161mm}
   \deltaabs=1pt \deltaref=1.15pt \deltabase=0pt }
\def\langformat
  {\addtolength{\topmargin}{-26mm}
   \setlength{\textheight}{245mm}
   \addtolength{\oddsidemargin}{-12.5mm}
   \setlength{\textwidth}{157mm}
   \deltaabs=2pt \deltaref=1.25pt \deltabase=2pt }
\kurzformat 
\makeatletter
\new@fontshape{msb}{m}{n}{%
   <5>msbm5%
   <6>msbm6%
   <7>msbm7%
   <8>msbm8%
   <9>msbm9%
   <10>msbm10%
   <11>msbm10%
   <12>msbm10 at10.95pt%
   <14>msbm10 at14.4pt%
   <17>msbm10 at17.28pt%
   <20>msbm10 at20.74pt%
   <25>msbm10 at24.88pt}{}
\extra@def{msb}{}{}
\newmathalphabet*\bbl{msb}{m}{n}

\new@fontshape{euf}{m}{n}{%
   <5>eufm5%
   <6>eufm6%
   <7>eufm7%
   <8>eufm8%
   <9>eufm9%
   <10>eufm10%
   <11>eufm10 at10.95pt%
   <12>eufm10 at12pt%
   <14>eufm14%
   <17>eufm14 at17.28pt%
   <20>eufm14 at20.74pt%
   <25>eufm14 at24.88pt}{}
\extra@def{euf}{\hyphenchar#1\m@ne
       \@tempdima\fontdimen2#1%
       \fontdimen3#1.4\@tempdima
       \fontdimen4#1.3\@tempdima}{}%
\newmathalphabet*\got{euf}{m}{n}

\newif\if@fewtab\@fewtabtrue
\makeatother
\begin{document}
\newcommand{\ncd}{\newcommand} \newcommand{\rcd}{\renewcommand}
\ncd{\be}{\begin{equation}}        \ncd{\ee}{\end{equation}}
\ncd{\bq}{\begin{quote}}           \ncd{\eq}{\end{quote}}
\ncd{\barr}{\begin{array}}         \ncd{\earr}{\end{array}}
\ncd{\sectreset}[1]{\section{#1}\setcounter{equation}{0}
                             \message{\thesection : #1}}
\ncd{\subsect}[1]{\vspace{4mm}\noindent{\bf #1}\\[2mm]\nobreak\noindent}
\rcd{\theequation}{\thesection.\arabic{equation}}
\ncd{\sps}{{super\-selec\-tion}}   \ncd{\st}{{sta\-tistics}}
\ncd{\rep}{{repre\-sen\-tation}}   \ncd{\vrep}{{vacuum \rep}}
\ncd{\emor}{{endo\-mor\-phism}}    \ncd{\cmor}{{canonical \emor}}
\ncd{\imor}{{iso\-mor\-phism}}     \ncd{\amor}{{auto\-mor\-phism}}
\ncd{\lc}{{light-cone}}            \ncd{\ea}{{exchange algebra}}
\ncd{\ca}{{current algebra}}       \ncd{\itw}{{inter\-twiner}}
\ncd{\rfb}{{reduced field bundle}} \ncd{\trafo}{{trans\-for\-ma\-tion}}
\ncd{\qft}{{quantum field theory}} \ncd{\cex}{conditional expectation}
\ncd{\irrep}{{irreducible \rep}}   \ncd{\emt}{stress-energy tensor}
\ncd{\pipo}{Pimsner-Popa bound}    \ncd{\vna}{von~Neumann algebra}
\ncd{\comrel}{{commutation relation}}    \ncd{\vn}{von~Neumann}
\ncd{\CGC}{{Clebsch-Gordan coefficient}} \ncd{\hpf}{hyperfinite}
\ncd{\cfb}{{conformal block}}
\rcd{\aa}{{\cal A}}     \ncd{\ii}{{\cal J}}    \ncd{\ff}{{\cal F}}
\ncd{\oo}{{\cal O}}     \ncd{\bb}{{\cal B}}    \ncd{\hh}{{\cal H}}
\ncd{\ch}{_{\rm ch}}    \ncd{\wt}{\widetilde}  \ncd{\ovl}{\overline}
\ncd{\two}{^{(2)}}      \ncd{\inv}{^{-1}}      \ncd{\sch}{^{(s)}}
\rcd{\a}{\alpha}  \rcd{\b}{\beta}  \ncd{\g}{\gamma}  \rcd{\d}{\delta}
\ncd{\sig}{{\sigma}}    \ncd{\eps}{\varepsilon} \ncd{\lam}{{\lambda}}
\rcd{\rho}{{\varrho}}   \ncd{\brho}{{\bar\varrho}}   \ncd{\po}{{\pi_0}}
\ncd{\bsig}{{\bar\sig}} \ncd{\vphi}{\varphi}
\ncd{\RR}{{\bbl R}}   \ncd{\NN}{{\bbl N}}   \ncd{\CC}{{\bbl C}}
\ncd{\ZZ}{{\bbl Z}}   \ncd{\DD}{{\wt D}}    \ncd{\BBB}{{\got B}}
\ncd{\III}{I\!I\!I}   \ncd{\II}{I\!I}
\ncd{\eins}{{\mathchoice {\rm 1\mskip-4mu l} {\rm 1\mskip-4mu l}
                         {\rm 1\mskip-4.5mu l} {\rm 1\mskip-5mu l}}}
\ncd{\LO}{\langle \Omega} \ncd{\RO}{\Omega \rangle}
\ncd{\sub}[2]{#1 \subset #2}
\ncd{\qbin}[2]{\left[{#1 \atop #2} \right]}
\ncd{\comp}{\makebox{\raise1pt\hbox{$\scriptstyle \circ$}}}
\ncd{\arr}{\! \rightarrow \!}
\ncd{\map}[2]{\colon #1 \arr #2} \ncd{\id}{{id}}
\parindent0mm \parskip2mm
\addtolength{\baselineskip}{\deltabase}
\setcounter{section}{0} \setcounter{equation}{0}
DESY 94-164 \hspace*\fill September 1994 \\
ESI 132 (1994)\hspace*\fill hep-th/9409165\\
 \vspace{4mm}
\rcd{\thefootnote}{\fnsymbol{footnote}}
\begin{center} {\Large \bf Characterizing invariants  \\[2mm]
for local extensions of \ca s} \\[9mm]
{\sc Karl-Henning Rehren} \\[3mm] II.\ Institut f\"ur Theoretische
Physik, Universit\"at Hamburg, \\ D-22761 Hamburg ({\sc Germany})
\\[5mm]
{\sc Yassen S.\ Stanev} and {\sc Ivan T.\ Todorov} \\[3mm]
Erwin Schr\"odinger International Institute for Mathematical Physics, \\
A-1090 Wien ({\sc Austria}) \\ and \\
Institute for Nuclear Research and Nuclear Energy, \\
Bulgarian Academy of Sciences, BG-1784 Sofia ({\sc Bulgaria})
\footnote{permanent address} \end{center}
\rcd{\thefootnote}{\arabic{footnote}}
\rcd{\Large}{\large}
\addtolength{\baselineskip}{-\deltaabs}
\bq {\bf Abstract:}
Pairs $\aa \subset \bb$ of local quantum field theories are studied,
where $\aa$ is a chiral conformal \qft\ and $\bb$ is a local extension,
either chiral or two-dimensional. The local correlation functions of
fields from $\bb$ have an expansion with
respect to $\aa$ into \cfb s, which are non-local in general. Two
methods of computing characteristic invariant ratios of
structure constants in these expansions are compared: $(a)$ by
constructing the monodromy \rep\ of the braid group in the space of
solutions of the Knizhnik-Zamolodchikov differential equation, and $(b)$
by an analysis of the local subfactors associated with the extension
with methods from operator algebra (Jones theory) and algebraic quantum
field theory. Both approaches apply also to the reverse problem: the
characterization and (in principle) classification of local extensions
of a given theory. \eq
\addtolength{\baselineskip}{\deltaabs}
\sectreset{Introduction}
The relevance of V.\ Jones' theory of (\vn) subfactors \cite{Jo} for
2-dimensional (2D) models of critical behaviour was first recognized in
the work of V.\ Pasquier on lattice models labelled by Dynkin diagrams
\cite{Pa}. A spectacular by-product of this realization was the ensuing
$ADE$ classification of $su(2)$ \ca\ models and minimal conformal
theories \cite{CIZ}. The above parallel was understood within the
Haag-Kastler algebraic approach to local \qft\ \cite{HK} in terms of the
Doplicher-Haag-Roberts (DHR) theory of \sps\ sectors and particle
statistics \cite{DHR} applied to chiral algebras \cite{FRSI,FRSII}, and
provided an explanation for the Jones index as a measure for the
violation of Haag duality (maximality of local observables) in a given
\rep, and relating it numerically to the statistical dimension
\cite{Lo}.

In the cited work on subfactors in \qft, the emphasis for the use of
the theory of subfactors was its application to individual \sps\
sectors of a given local theory, and the derivation of invariant
`charge quantum numbers' such as statistical dimensions and Markov
traces. In contrast, here we shall consider a pair of local theories,
one extending the other, as a subfactor (actually, a net of local
subfactors) and apply the properly adapted Jones theory to describe the
`position' of the subtheory in its extension. This point of view opens
the way to a detailed understanding of the behaviour of \sps\ sectors
when one passes from one theory to the other by a generalized Mackey
induction and restriction prescription \cite{FLRR}.

In particular, given that the position of a subtheory in another
theory is encoded and characterized by a subfactor, then subfactor
theoretical methods can be applied to conformal models and their local
extensions, and must give detailed answers comparable with the $ADE$
classification and related results obtained by conventional methods of
conformal \qft.

The present article is a comparative study of conventional field
theoretical methods on the one hand and the theory of subfactors on
the other hand in application to the same problem: local extensions of
local quantum field theories. A local extension is determined by the
correlation functions of the extending fields. In chiral \ca\ models of
conformal field theory, the extending fields necessarily correspond to
primary fields of the unextended theory with bosonic, i.e., integer
conformal dimension $\Delta$. Their 4-point functions are linear (for
chiral extensions) or bilinear (for 2D extensions) combinations of
\cfb\ functions which are monodromy free inspite of the non-trivial
braid group transformation of the individual \cfb s. Moreover, unlike
the chiral vertex operators of the unextended theory whose fusion
rules coincide with the intrinsic composition law of \sps\ charges
provided by the DHR theory, the extending local fields must satisfy
truncated fusion rules which involve only other bosonic fields, and
which are therefore only majorized by the DHR fusion.

Both the truncated fusion rules and the ratios of structure constants
(amplitudes of \cfb\ functions) in the said combinations are
characteristic quantities for a pair of a chiral \ca\ and its extension.
They are computed by both methods. In the first part of the article
(Sects.\ 2 and 3), we study the monodromy behaviour of the solutions of
the Knizhnik-Zamolodchikov (KZ) equation and compute the braid invariant
quadratic forms which determine the local 4-point functions of the
two-dimensional extensions. Apart from the generic two-dimensional
extension (corresponding to the $A$ series of the $ADE$ classification),
and the chiral $D$ series extensions which correspond to a global
$\ZZ_2$ symmetry, we concentrate on the exceptional chiral $E_6$ and
$E_8$ extensions of $su(2)$ \ca s. We compute explicitly the relative
amplitudes of the $A$ and $E$ theories, which turn out to be rational
numbers. In the second part (Sects.\ 4 and 5), we study the position of
the operator algebra of the unextended subtheory within its extension,
in terms of the theory of subfactors. Remarkably, the relevant
information already resides in a single pair of local \vna s. We
analyze which quantities in the general theory of subfactors, when
applied to a given local field extension, contain the desired
information about the truncated fusion rules and the relevant ratios of
structure constants. We describe how to compute these data in terms of
the subtheory (interpreted as the physical observables) and its \sps\
structure.

While the first method will be easier to use in specific models and as
long as one is interested only in 4-point functions, the second method
is part of a general theory of local field extensions, confined neither
to two dimensions nor to conformal quantum field theories. It covers
also the standard situation of four-dimensional theories with a compact
gauge group. (In this latter case, the method essentially reduces to
harmonic analysis and partial wave expansions based on the \rep\ theory
and \CGC s of the gauge group.) It has the advantage to treat all
$n$-point functions at one stroke. However, in practice it requires to
solve in a first step a complicated non-linear system for the
`generalized \CGC s', which we have carried out only for the simplest
model of a field extension which is not due to a gauge group.

The local extensions of chiral $su(2)$ \ca s studied in Sects.\ 2 and 3
are distinguished to have the same \emt\ as the original theory, the
\emt\ implicitly entering the analysis through the KZ equations. If the
extending fields are currents of dimension $\Delta=1$, this means that
the extension is a `conformal embedding' \cite{BNSW}. On the other
hand, in Sects.\ 4 and 5, we assume the index of the inclusion to be
finite. Indeed, for pairs of chiral \ca s, these two selection criteria
are equivalent. Namely, both the finiteness of the index and the
triviality of the coset \emt\ are equivalent to the finiteness of the
branching of the \vrep\ of the extended theory upon restriction to the
subtheory.

Let $\aa \subset \bb$ be a conformal embedding \cite{BNSW} of two
chiral quantum field theories like the \ca s $\aa_{10}(A_1) \subset
\aa_1(B_2)$ where $A_1 = su(2)$ and $B_2 = spin(5) \simeq sp(4)$ refer
to the Lie algebras underlying the \ca s, and the subscripts refer to
the level $10$ resp.\ $1$ of the central extension. The embedding gives
rise to a pair of braid-invariant quadratic forms $M$ and $\wt M$ in the
space of 4-point \cfb s of the subtheory $\aa$ with four given external
quantum numbers (\sps\ charges) such as isospins $I \leq k/2$ for $\aa =
\aa_k(A_1)$. The quadratic forms serve to express 2D correlation
functions in terms of chiral \cfb s, and turn out to completely
characterize the model. The form $M$ corresponds to the `diagonal' WZNW
theory \cite{WZW} over $\aa$, i.e., to the $A_{k+1}$ theory in the
$ADE$ classification of $su(2)$ \ca\ models at level $k$ \cite{CIZ}.
The eigenvalues $D_\lam^{(k,I)}$ of $M$, in the case of 4 equal
external isospins $I$, are the squares of the structure constants
\be D_\lam^{(k,I)} \equiv N_{II\lam}^2 \qquad (\lam = 0,1,\dots
\min(2I,k-2I) \equiv m_{kI}) \ee
for the $s$-channel fusion of two of the isospin $I$ charges into
isospin $\lam$ intermediate states. We recall that for $4I>k$, the
subspace of 4-point blocks with $\lam>m_{kI}$ corresponds to
`unphysical' correlations which violate positivity. Only the `physical'
blocks contribute to $M$ and $\wt M$.

The form $\wt M$ corresponds to the diagonal theory over the chiral
extension $\bb$. Since the local fields of the latter are in general
non-diagonal with respect to $\aa$, the form $\wt M$ is a non-diagonal
matrix in the $s$-channel basis of \cfb\ functions which diagonalizes
$M$. The ratios of the diagonal elements of the form $\wt M$ to the
corresponding eigenvalues (1.1) of $M$ are invariant under rescaling of
the 4-point blocks and thus provide a basis-independent characteristics
of the non-diagonal theory associated with the form $\wt M$. Such ratios
were already considered in the above-mentioned pioneer work by Pasquier
\cite{Pa}, and have later been computed for specific conformal
embeddings \cite{FGPFKS}. We shall provide in Sect.\ 3 below an
independent computation using previous work on monodromy \rep s of the
braid group \cite{STHI,STHII}.

Let us turn to the subfactor point of view. As we shall see, one can
characterize a local field extension $\bb$ of a given theory $\aa$ in
terms of a triple $(\rho,W,X)$. Here $\rho$ is a localized \emor\ of
$\aa$ equivalent to a reducible \rep\ $\pi$ of $\aa$ (the restriction
of the \vrep\ of $\bb$), $W$ is an isometric observable (i.e., $W^*W =
\eins$) such that $E = WW^*$ projects onto the \vrep\ $\po$ of $\aa$
contained in $\pi$, and $X$ is a second isometry satisfying a system of
identities with $W$, involving $\rho$, which guarantees the possibility
to recover the local extension from these data. The states in the
non-trivial subsectors of $\pi$ are created from the vacuum by the
extending fields. The operator $X$ may be considered as a generating
functional for all the relevant `generalized \CGC s' associated with
the inclusion. The mathematical concept behind this notion is a
`harmonic analysis' for subfactors, which generalizes the ordinary
harmonic analysis in the case of a compact gauge symmetry. The
coefficients determine both the truncated operator product expansions
and the amplitudes of `partial waves' in correlation functions of local
charged fields. These partial waves of the subfactor harmonic analysis
will be identified with the \cfb s in chiral \ca\ models, and the \CGC
s coincide with the structure constants entering the quadratic forms as
discussed before.

It is important to note that also in this general context, there is
always a `standard' extension (corresponding to the generic
braid-invariant quadratic form $M$ in the case of chiral \ca s) which
can be used to fix the normalizations, i.e., to absorb the uncontrolled
kinematical model characteristics, by computing invariant double ratios
of amplitudes.

Our article is organized as follows. We review in Sect.\ 2 the monodromy
\rep\ of the mapping class group $\BBB_4$ of the 2-sphere with 4
punctures in the space of solutions of the KZ equation, and write down
the generic braid invariant form corresponding to the $A$ series in the
$ADE$ classification. In Sect.\ 3, the explicit computations are done
for two models of special interest, the $E_{\rm even}$ series conformal
embeddings labelled $E_6$ and $E_8$.

In Sect.\ 4, we turn to the theory of subfactors (of finite index) and
introduce some of the basic concepts which are of particular relevance
for the application to (local) field extensions. In Sect.\ 5, the
connection with chiral vertex operators is established, and the general
method to compute relative structure constants in terms of subfactors is
presented. The method is then applied to the $E_6$ inclusion and
reproduces the results obtained in Sect.\ 3.

The two parts consisting of Sects.\ 2,3 and Sects.\ 4,5, respectively,
are to a large extent independent of each other. The reader may start
with either part according to personal preference. Our point is the
comparison of the conceptually different guises under which the same
quantities arise in the two approaches.

\sectreset{Braid invariant positive forms in the space of 4-point
blocks}
We start with the algebra of observables $\aa_k = \aa_k(A_1)$ generated
by the level $k$ chiral $su(2)$ currents. It includes the chiral
Sugawara \emt. The primary chiral vertex operators $V_I$ \cite{TK} which
intertwine the vacuum sector with the \sps\ sector of charge $I$ (=
positive energy \rep\ of $\aa_k$ with lowest energy eigenstates of
isospin $I$) are assumed to have homogeneous local \comrel s with the
currents (`local gauge covariance') and with the \emt\ tensor
(`reparametrization covariance'). These assumptions imply the KZ
equation \cite{KZ} as well as the relation between isospin and conformal
(scaling) dimension $\Delta_I$
\be (k+2)\Delta_I = I(I+1) \qquad (2I = 0,1,\dots k). \ee

\subsect{2A. The mapping class group and its monodromy \rep s}
We consider 4-point functions for four primary fields of isospin $I$. We
first construct the $2I+1$ dimensional \rep\ of the mapping class group
$\BBB_4$ of the 2-sphere with 4 punctures acting in the
$(2I+1)$-dimensional space of all 4-point solutions of the
corresponding KZ equation, into which the level $k$ enters only via
the complex phase
\be q = \exp\left(\frac{i\pi}{k+2}\right). \ee
Unless $k$ is a positive integer, this space of solutions violates the
positivity of correlation functions, and the \rep\ of $\BBB_4$ is not
unitarizable. Yet, it is computationally advantageous to deal with
generic $q$ in a first step. At a given level $k \in \NN$, positivity is
still violated for $4I>k$, and one has therefore, in a second step, to
restrict to the $(m_{kI}+1)$-dimensional invariant `physical' subspace
spanned by the $s$-channel blocks $s_\lam^{(I)}$ with $\lam$ in the
range of (1.1).

The (projectively represented) mapping class group $\BBB_4$ can be
identified as the braid group of 4 strands on the sphere with
generators $B_i$, $i=1,2,3$, such that
\be B_1B_3 = B_3B_1\;, \qquad\qquad B_iB_{i+1}B_i = B_{i+1}B_iB_{i+1}
\quad (i=1,2) \ee
\be B_1B_2B_3^2B_2B_1 = B_3B_2B_1^2B_2B_3 = q^{-4I(I+1)} \ee
satisfying the additional relation
\be (B_1B_2B_3)^4 = q^{-8I(I+1)}. \ee
(In the standard definition of $\BBB_4$, the relations (2.4) and (2.5)
are assumed to hold with $q=1$; here we are dealing with a projective
\rep, or equivalently, with a central extension of the mapping class
group.) It can be proven, using only the above relations, that the
monodromy operators $B_1^2$ and $B_3^2$ are equal. It then follows
from (2.4) that the `fusion' matrix $F$ has square 1:
\be B_1B_2B_1 \equiv B_2B_1B_2 =: (-1)^{2I}q^{-2I(I+1)}F,
\qquad F^2 = \eins. \ee
$F$ plays the role of a $6j$ symbol (in general, for 4-point blocks
of different isospins $I_i$, its matrix elements require 6 labels
$F_{\lam\mu} = F^{I_1I_2I_3I_4}_{\lam\mu}$).

An analysis of the solutions of the KZ equation shows that (in the
case at hand with four equal isospins $I$), actually the generators
$B_1$ and $B_3$ coincide:
\be B_1 = B_3. \ee
Moreover, there exists a basis of solutions \cite{STHI} for which the
fusion matrix has only non-zero elements on the second diagonal,
\be F_{\lam\mu} = \d_{\lam+\mu,2I} \qquad (\lam,\mu = 0,1,\ldots 2I),
\ee
while $B_1$ is upper triangular:
\be (B_1)_{\lam\mu} = (-1)^{2I-\mu}q^{\mu(\lam+1)-2I(I+1)}
\qbin{2I-\lam} {\mu-\lam}. \ee
Here, $\qbin nm $ are the (real) $q$-binomial coefficients vanishing for
$n<m$ and otherwise given by
\be \qbin nm = \frac{[n]!}{[m]![n-m]!}, \qquad [n]! = [n][n-1]!, \quad
[0]! = 1, \ee
\be [n] = \frac{q^n-q^{-n}}{q-q\inv} =
\frac{\sin\frac{n\pi}{k+2}}{\sin\frac{\pi}{k+2}}. \ee
We are using a non-unitary basis (even for $4I \leq k$ when $B_1$ is
unitarizable) which has the following advantages: \\
$(i)$ it exhibits no singularities for $4I \geq k+2$ ($2I \leq k$, $q$
given by (2.2)); \\
$(ii)$ the entries of the braid matrices and of the invariant forms are
elements of the cyclotomic field ${\bbl Q}(q^{1/2})$ (or ${\bbl Q}(q)$
for integer $I$; $q^{k+2} = -1$). \\
We anticipate here, that the ratios of structure constants we are
finally interested in (eqs.\ (3.8), (3.9), and (3.15) below) turn out
rational and are therefore invariant under Galois \amor s $q \mapsto
q^n$ ($n$ and $2k+4$ coprime) of this field.

The second generator, $B_2$, of $\BBB_4$ is a conjugate to $B_1$ by $F$:
\be B_2 = FB_1F \qquad (B_1 = FB_2F) \ee
and appears as a lower triangular matrix.

It is noteworthy that this monodromy \rep\ of $\BBB_4$ can in fact be
derived without a detailed study of the solutions of the KZ equations.
Indeed, the eigenvalues of $B_1$ are already read off the 3-point block
functions, which are just powers of the coordinate differences. In a
basis in which the fusion matrix $F$ has the form (2.8) and $B_1$ is
upper triangular, the non-diagonal entries of $B_1$ and the matrix
$B_2$ are determined by (2.6) up to a rescaling of the basis. As it was
already noted, the ratios of interest will turn out to be invariant
under such a rescaling, too.

\subsect{2B. The generic $\BBB_4$ invariant symmetric form}
The local 4-point function of the two-dimensional theory is defined by
a hermitian braid invariant form $M$ in the space of 4-point blocks:
\be \langle \Phi_I \Phi_I \Phi_I \Phi_I \rangle \propto G_4 =
\sum_{\lam\mu} \bar f_\lam M_{\lam\mu} f_\mu \quad {\rm with} \quad
M^+ = M = B^+MB \qquad (B \in \BBB_4) \ee
where an appropriate power of the coordinate differences has been split
off as usual, and $f$ resp.\ $\bar f$ depend only on the conformally
invariant cross ratios of coordinate differences on the left- resp.\
right-moving \lc. (For further details on the choice of basis $f_\lam$
see \cite{STHI}.)

The above non-unitary realization of $B_i$ has the advantage that the
inverse generators are just given by the complex conjugate matrices
\be B_i\inv = \ovl B_i \qquad {\rm since} \quad \bar q = q\inv.\ee
The same is trivially true for $F$.

We are thus looking for a real symmetric form $M = (M_{\lam\mu})
= {^tM}$ satisfying the braid invariance condition
\be ^tB_iM = MB_i \qquad (i=1,2). \ee
\bq
{\bf Proposition 2.1:} \cite{STHII} For every $q \neq 0$ there exists a
diagonalizable $\BBB_4$ invariant symmetric form in the space of 4-point
solutions of the KZ equation with four isospins $I$
\be M = {^tS}DS \qquad {\rm where} \quad D_{\lam\mu} = D_\lam
\d_{\lam\mu}. \ee
At the values $q=e^{\frac{i\pi}{k+2}}$ ($k \in \NN$), the diagonal
matrix $D$ has $m_{kI}+1$ non-zero elements (with $m_{kI}$ given by
(1.1)):
\be D_\lam \equiv D_\lam^{(k,I)} = \left\{\frac{[\lam]![2I+1+\lam]!}
{[2I+1]![2\lam]!}\right\}^2 \frac 1{[2\lam+1]}, \qquad (\lam =
0,\ldots m_{kI}). \ee
If $4I>k$, then $D_\lam$ vanish for $m_{kI} < \lam \leq 2I$. The
transformation matrix $S$ is a real upper triangular matrix with
elements
\be S_{\lam\mu} = (-1)^{\mu-\lam} \qbin\mu\lam
\frac{[2I-\lam]![2\lam+1]!}{[2I-\mu]![\lam+\mu+1]!} \quad
{\rm for} \quad 0 \leq \lam \leq \mu \leq m_{kI} \ee
and $S_{\lam\mu} = \d_{\lam\mu}$ for $\lam>m_{kI}$.
\eq
{\it Sketch of a proof:} We consider the similarity transformation
\be B \mapsto B\sch := SBS\inv. \ee
The specific block form $S = \pmatrix{\Sigma & \Sigma' \cr 0 & \eins
\cr}$ -- where $\Sigma$ is given by (2.18) and the rectangular block
$\Sigma'$ is only present when $4I>k$ -- implies the block form of the
inverse matrix $S\inv = \pmatrix{\Sigma\inv & -\Sigma\inv\Sigma' \cr 0
& \eins \cr}$ with
\be S\inv_{\lam\mu} = \Sigma\inv_{\lam\mu} = \qbin\mu\lam
\frac{[2I-\lam]![\lam+\mu]!}{[2I-\mu]![2\mu]!} \qquad{\rm for} \quad
0 \leq \lam \leq \mu \leq m_{kI}. \ee
The transformation (2.19) brings $B_1$ in a reduced form for $4I>k$ and
diagonalizes it for $4I \leq k$; in both cases
\be (B_1\sch)_{\lam\mu} = \d_{\lam\mu} (-1)^{2I-\lam}
q^{\lam(\lam+1)-2I(I+1)} \qquad {\rm for} \quad \lam,\mu \leq m_{kI}.
\ee
In particular, the basis $s_\lam = S_{\lam\mu}f_\mu$ of \cfb s has
definite $B_1$ mono\-dromy on the physical subspace $0 \leq\lam\leq
m_{kI}$. (For this reason we call $s_\lam$ the $s$-channel basis.)

It follows that $B_1\sch$ commutes with $D$ and hence (2.15) holds
for $i=1$. Verification of invariance of $M$ with respect to $B_2$ or
$F$ requires more work. One should either use the explicit form of $M$:
\be M_{\lam\mu} =
\frac{(-1)^{\lam+\mu}[\lam]![\mu]!}{[2I-\lam]![2I-\mu]![2I+1]!^2}
\sum_{\nu=0}^{m_{kI}} \frac{[2I+\nu+1]!^2[2I-\nu]!^2[2\nu+1]}
{[\lam+\nu+1]![\mu+\nu+1]![\lam-\nu]![\mu-\nu]!} \ee
or transform $F$ to the $s$-channel basis ($F \mapsto F\sch =
SFS\inv$) -- see below.

{\it Remarks:} $\triangleright$ An expression of the type (2.16),
(2.22) for the invariant form was first derived in
\cite[Sect.\ 6]{STHII} using quantum group techniques. The present
formulae differ slightly because of a different normalization of the
basis. They are related by $[2I+1]^2 M_{\lam\mu} = \qbin{2I}\lam
\qbin{2I}\mu Z_{\lam\mu}$. Such a change of basis does not affect the
ratios of structure constants to be computed below. \\
$\triangleright$ The Proposition explicitly provides the transition
matrix to the $s$-channel basis, from which, together with the spectrum
(2.21) of the braid matrix, all the basis-independent quantities of
interest in the sequel will be obtained by direct computations.

The braid invariant 2D 4-point function now assumes a diagonal form
in the physical $s$-channel basis $s_\lam$ with $\lam \leq m_{kI}$
\be G_4 = \sum_{\lam=0}^{m_{kI}} D_\lam^{(k,I)} \bar s_\lam s_\lam. \ee

Summing up we see that, at the quantized values (2.2) of $q$, and more
generally for any $q$ such that $q^{k+2} = -1$, the $(2I+1)$-dimensional
\rep\ $\BBB_4$ of the mapping class group is reducible when $4I>k$. It
is also non-unitarizable, the generators $B_i$ being not diagonalizable
(for $4I \geq k+2$). It is the kernel of the form $M$ that carries a
non-unitary factor \rep. The $(m_{kI}+1)$-dimensional sub\rep\
$\BBB_4^{(k,I)}$ preserves a
non-degenerate positive form (2.23) and is hence unitarizable. The
resulting $(m_{kI}+1)$-dimensional \rep\ may, in general, still be
reducible. As we shall see in Sect.\ 3A., this fact is responsible for
the possible existence of non-diagonal local extensions.

The $s$-channel reflection matrix $F\sch$ (which is related to the
exchange of the factors 1 and 3 in (2.13) and which,
for four generic isospins, encodes the entire fusion information of the
model) is, not surprisingly, considerably more complicated than the
original expression (2.8). We have computed it from
\[ F\sch = SFS\inv = SU\inv = US\inv \]
in terms of the above $s$-channel transition matrix $S$ which
diagonalizes $B_1$, and the $u$-channel transition matrix $U = SF$
which diagonalizes $B_2$:
\[ U_{\lam\mu} = S_{\lam,2I-\mu} = (-1)^{2I-\lam-\mu}
\qbin{2I-\mu}\lam \frac{[2\lam+1]![2I-\lam]!}{[\mu]![2I+\lam-\mu+1]!} \]
giving
\be F\sch_{\lam\mu} =
\frac{[\mu]![2\lam+1]![2I-\lam]!}{[\lam]![2\mu]![2I-\mu]!}
\sum_{\nu=0}^\mu \frac{(-1)^{2I-\lam+\nu}[\mu+\nu]![2I-\nu]!^2}
{[\nu]!^2[\mu-\nu]![2I-\lam-\nu]![2I+\lam-\nu+1]!} . \ee
We note that, even if we use expressions (2.18) and (2.20) beyond the
range of their validity (i.e., for $\mu > m_{kI}$ when $4I>k$) where
some of the entries of the transition matrix $S$ and $S\inv$ are ill
defined at the value (2.2) of $q$, the $F$ matrix (2.24) is finite in
the physical range $0 \leq \lam,\mu \leq m_{kI}$. Moreover, the
restricted $(m_{kI}+1) \times (m_{kI}+1)$ matrices $B_1\sch$, $F\sch$,
and
\be B_2\sch = F\sch B_1\sch F\sch \qquad (F^{(s)2} = \eins) \ee
still satisfy the relations (2.3) -- (2.7). This is a non-trivial
statement for $4I>k$.

The braid invariance of the two-dimensional Green's function (2.23)
implies the relation
\be F\sch_{\lam\mu} D^{(k,I)}_\lam = D^{(k,I)}_\mu F\sch_{\mu\lam}
\ee
with the positive eigenvalues $D$ of the form $M$ given by (2.17). Hence
on the one hand, the $s$-channel $F$ matrix is symmetrizable, and on the
other hand, the ratios of amplitudes for the diagonal extension are
given by
\be \frac{N_\lam^2}{N_\mu^2} \equiv \frac{D_\lam}{D_\mu} =
\frac{F\sch_{\mu\lam}}{F\sch_{\lam\mu}}. \ee

\sectreset{Ratios of structure constants for the $E_6$ and the $E_8$
models}
The braid-invariant 4-point functions (2.13), (2.23) give the monodromy
free Green's functions for the 2D local extensions of the chiral $su(2)$
\ca s $\aa_k$ corresponding to the $A_{k+1}$ series in the $ADE$
classification.

There exists an infinite set of extensions of the $su(2)$ \ca s for
level $k$ a multiple of 4, corresponding to the $D_{2n}$ series ($2n =
k/2+2$). In these models, the chiral algebras are extended by an
$\aa_k$-primary simple current: a Bose field of isospin and conformal
dimension
\be I=\frac k2 \qquad {\rm and} \qquad \Delta_I =
\frac{I(I+1)}{k+2} = \frac k4 \in \NN. \ee
The inclusion of the (nets of) algebras $\aa_k$ in the resulting field
algebras are well understood: it is of the DHR type \cite{DHR,DR}
with a global $\ZZ_2$ gauge group which singles out the `observables'
$\aa_k$ as the gauge invariant elements \cite{FGK} (for a recent review
and further references see \cite{IT}).

Here we shall deal with the more interesting exceptional extensions
corresponding to conformal embeddings \cite{BNSW}. These are not of
the DHR type, i.e., the $\aa_k$ subalgebras are not the gauge invariants
with respect to some global gauge group.

\subsect{3A. Pairs of braid invariant quadratic forms for exceptional
embeddings}
There are just two non-trivial chiral extensions of $\aa_k(A_1)$
corresponding to the conformal embeddings
\[ \aa_{10} = \aa_{10}(A_1) \subset \aa_1(B_2) = \bb_{10} \qquad\qquad
(E_6) \]
\[ \aa_{28} = \aa_{28}(A_1) \subset \aa_1(G_2) = \bb_{28} \qquad\qquad
(E_8) \]
where the labels $E_6$ and $E_8$ refer to the $E$ series of the $ADE$
classification \cite{CIZ}. The \sps\ structure of the observables
in the `diagonal' \rep\ space of the respective field extensions is
encoded in the exceptional partition functions

$ \quad \hfill Z(E_6) = |\chi_1+\chi_7|^2 + |\chi_4+\chi_8|^2 +
|\chi_5+\chi_{11}|^2 \hfill\qquad (3.2a)$

$ \quad \hfill Z(E_8) = |\chi_1+\chi_{11} +\chi_{19} +\chi_{29}|^2 +
|\chi_7+\chi_{13} + \chi_{17}+\chi_{23}|^2 \hfill (3.2b) $

where the subscripts on the modular characters $\chi$ stand for the
dimensions, $2I+1$, of the $SU(2)$ \rep s labelling the \sps\ sectors of
$\aa_k$. Every term in these sums corresponds to a \sps\ sector of the
extended chiral \ca\ $\bb$, and every sum of modular characters
appearing in each term determines the branching of the corresponding
sector upon restriction to $\aa_k$. In particular, the first term added
to the vacuum character $\chi_1$ in (3.2) corresponds to the $\Delta_I =
1$ sector of $\aa_k$ generated by the $\bb_k$ currents orthogonal to the
$\aa_k$ currents. These are the (7 component) $I=3$ primary fields for
the $\aa_{10}$ theory in the $E_6$ case, and the (11 component) $I=5$
primary fields for the $\aa_{28}$ theory in the $E_8$ case.

The fact that an $\aa_k$-primary field $\phi_I$ (with integer
dimension $\Delta_I$) is a local Bose field in the extended $\bb_k$
theory means that, in particular, there exists a braid invariant
linear combination of 4-point blocks of the associated chiral vertex
operators. Namely, the commutation of two fields $\phi_I$ corresponds
to a monodromy operation on the \cfb\ functions. In other
words, the \rep\ $\BBB_4^{(k,I)}$ must be reducible and have an
invariant subspace of joint eigenvectors of $B_i$ with eigenvalue 1.

In the $s$-channel basis of eq.\ (2.23), these are combinations of the
form

$\quad\hfill \qquad E_6\quad (k=10): \qquad\qquad s_0^{(3)} +
\wt D_{03} s_3^{(3)} \qquad\qquad \hfill (3.3a)$

$\quad\hfill E_8\quad (k=28): \qquad s_0^{(5)} + \wt D_{05}
s_5^{(5)} + \wt D_{09} s_9^{(5)} \hfill (3.3b)$

\setcounter{equation}{3}
where $\wt D_{\lam\mu} \equiv \wt D_{\lam\mu}^{(k,I)}$ depend on the
model, and $\wt D_{00} = 1$ is chosen as a normalization.
Two-dimensional correlation functions then result as products of two
chiral functions (3.3), one for either chiral \lc. They are thus
bilinear in $(\bar s_\lam,s_\mu)$ corresponding to a non-diagonal
version of (2.23) with $D$ replaced by $\DD$ where
\be \DD_{\lam\mu} = \DD_{0\lam}\DD_{0\mu} = \DD_{\mu\lam}. \ee

The expressions (3.3) are $\BBB_4^{(k,I)}$-invariant. $B_1$-invariance
is automatic since all $s$-channel functions $s^{(I)}_\lam$ contributing
to (3.3) correspond to the same $B_1$ eigenvalue $1 (=-q^{k+2})$, and it
excludes by the same argument all other $s$-channel contributions with
$\lam$ different from 0 or 3 ($E_6$) resp.\ 0,5,9, or 14 ($E_8$). The
non-zero elements of $\DD$ are determined from $F\sch$ invariance:
$^tF\sch\DD = \DD F\sch$. It is sufficient to use the equation
\be  (^tF\DD)_{0\mu} = (\DD F)_{0\mu} = 0
\qquad {\rm for} \quad \mu = 1,2. \ee
This gives for the isospin $I=3$ current in the $k=10$ model:
\be \DD_{03} = -\frac{F\sch_{01}}{F\sch_{31}}
= -\frac 1{[5]} = -\frac 1{2+\sqrt3} \qquad (k=10, I=3) \ee
and for the isospin $I=5$ current in the $k=28$ model:
\be \DD_{05} = \frac{F\sch_{02}F\sch_{91}-F\sch_{01}F\sch_{92}}
{F\sch_{51}F\sch_{92}-F\sch_{52}F\sch_{91}}, \qquad
\DD_{09} = \frac{F\sch_{01}F\sch_{52}-F\sch_{02}F\sch_{51}}
{F\sch_{51}F\sch_{92}-F\sch_{52}F\sch_{91}} \qquad
(k=28,I=5) \ee
which can be computed from (2.24).

We note that by a change of scale for the $s$-channel basis functions,
$D_{\lam\mu}$ and $\DD_{\lam\mu}$ change by the same factor, hence
their ratios are invariant under rescaling. It is remarkable that
these invariant ratios are found to be rational numbers:
\be \frac{\DD_{33}}{D_{33}} = 2 \qquad (k=10, I=3), \ee
\be \frac{\DD_{55}}{D_{55}} = \frac 94, \quad
\frac{\DD_{99}}{D_{99}} = \frac 54 \qquad (k=28, I=5). \ee
{\it Remark:} In a unitary basis in which $D_{\lam\mu} = \d_{\lam\mu}$,
the matrix $F\sch$ will become symmetric (and unitary) due to (2.26).
This unitarized $\widehat F$ can be obtained from our $F$ setting
\be \widehat F_{\lam\mu} = (sign\,F_{\lam\mu})
\sqrt{F_{\lam\mu}F_{\mu\lam}}. \ee
In such a unitary basis, the above ratios will coincide with
$\DD_{\lam\lam}$.

\subsect{3B. The braid group \rep\ in the Ramond sector}
The extended model $\bb_{10} = \aa_1(B_2)$ (see Sect.\ 3A.) is
parallel in many respects to the Ising model and the $su(2)$ level 2
\ca\ theory. All three models have three \sps\ sectors with
identical fusion rules, and involve a simple current of dimension
$\Delta = \frac 12$. For $\bb_{10}$, this field is the $SO(5)$ vector
field $\psi$ which is also an irreducible $\aa_{10}$ primary field of
isospin 2.

The state space of the fermionic field $\psi$ splits into two \irrep s
with respect to the extended `super \ca' generated by
$\psi(z)$: the Neveu-Schwarz sector $\hh_1 \oplus \hh_5$, and the
Ramond sector $\hh_2$, where $\hh_d$ denote the level 1 $spin(5)$
\ca\ \rep s labelled by the dimension $d$ of their lowest
energy subspace. The correlation functions of $\psi$ are single-valued
in the Neveu-Schwarz sector, and double valued in the Ramond sector.

Furthermore, in all three models, the primary dimension in the Ramond
sector is related to the Virasoro central charge
\be \Delta = \textstyle \frac 18 c, \ee
$c$ being given as $\frac 12$ times the number of components of
$\psi$ ($c = \frac 52$ for $\bb_{10}$).

We proceed to compute the $4 \times 4$ braid matrices in the
$s$-channel basis of all $\aa_{10}$ \cfb s of four fields of isospin
$I = \frac 32$ and dimension $\Delta = \frac 5{16}$ which belong to the
Ramond sector of $\bb_{10}$ (see eq.\ (3.2$a$)). Then we determine the
sub\rep\ acting in the subspace of \cfb s of the extended theory
$\bb_{10}$ which constitute the 2D local Ramond 4-point functions.

Applying (2.21) and (2.24) for $I = \frac 32$, we obtain
\be B_1\sch = q^{\frac 92}
\pmatrix{ 1&0&0&0 \cr 0& -q^2 &0&0 \cr 0&0& q^6 &0 \cr 0&0&0&1 \cr }
\qquad (k=10, q = e^{\frac{i\pi}{12}}, I=\textstyle \frac 32), \ee
and
\be F\sch = \pmatrix{
\frac{1-[3]}{3[2]} & \frac{[3]-1}3 & -\frac{[3]}{3[2]} & \frac 13 \cr
\frac{4-[3]}{3} & -\frac1{[2]} & 0 & \frac2{3[2]} \cr
-\frac{[3]}{2[2]} & 0 & \frac{[3]}{2[2]} & \frac{[3]-1}6 \cr
1 & \frac{[3]}{[2]} & \frac{[3]-1}3 & \frac{4-[3]}{6[2]} \cr }
 = \pmatrix{
\frac{1-\sqrt3}{\sqrt6} & \frac 1{\sqrt3} & - \frac{\sqrt2}3 & \frac 13
\cr \frac{\sqrt3-1}{\sqrt3} & \frac{1-\sqrt3}{\sqrt2} & 0 & \sqrt2
\frac{\sqrt3-1}3 \cr -\frac 1{\sqrt2} & 0 & \frac 1{\sqrt2} &
\frac 1{2\sqrt3} \cr 1 & \sqrt2 & \frac 1{\sqrt3} & \frac{2-\sqrt3}
{\sqrt6} \cr }.       \ee
The first matrix displayed here was computed with $q$-number identities
valid for every Galois transform of $q$. Evaluating $[3] = \sqrt2[2] =
1+\sqrt3$ at $q = e^{\frac{i\pi}{12}}$, one obtains the second matrix
(3.13).

We are now looking for an $E_6$-type braid invariant $s$-channel
quadratic form $\DD \equiv \DD^{(10,2)}$
\be \DD = \pmatrix{ 1 & 0 & 0 & \wt N_3 \cr
0 & 0 & 0 & 0 \cr 0 & 0 & \wt N_2^2 & 0 \cr
\wt N_3 & 0 & 0 & \wt N_3^2 \cr } \qquad{\rm where}\quad
\wt N_\lam = \frac {\wt N_{\frac 32 \frac 32 \lam}}
{\wt N_{\frac 32 \frac 32 0}}. \ee
The equality of the first and the last eigenvalue of $B_1\sch$ (eq.\
(3.12)) ensures $B_1$-invariance of $\DD$. The real parameters
$\wt N_\lam$ can be determined from $F$-invariance ${^tF}\wt M = \wt
MF$ of the quadratic form $\wt M = {^tS}\DD S$, which implies
\[ F\sch_{01}+F\sch_{31}\wt N_3 = 0, \qquad F\sch_{20}
\wt N_2^2 = F\sch_{02} + \wt N_3F\sch_{32}.\]
This yields $\wt N_3 = -1/\sqrt6$ and $\wt N_2^2 = 1$ for
$I=\frac 32$. We obtain the invariant ratios with the structure
constants $D_\lam \equiv  N_\lam^2$ of the diagonal theory given by
(2.17) or by (2.27):
\be \frac {\wt N_3^2}{N_3^2} =
\frac {F\sch_{10}F\sch_{01}}{F\sch_{13}F\sch_{31}} = \frac 12\,,
\qquad \frac {\wt N_2^2}{N_2^2} = 1 - \frac
{F\sch_{01}F\sch_{32}}{F\sch_{02}F\sch_{31}} = \frac 32
\qquad (k=10, I= \textstyle\frac 32). \ee
The same result is obtained for the invariant ratio of structure
constants for the isospin $I=\frac 72$ field, as expected since the
latter is the `partner' of the isospin $I=\frac 32$ field in the
partition function (3.2$a$), related by the simple current of isospin
$5$. Indeed, according to (2.1),
\[ \textstyle \Delta(\frac72) - \Delta(\frac32) = \frac{21}{16} -
   \frac{5}{16} = 1, \]
and hence the matrices $B_1\sch$ (projected into the physical
subspace of $s$-channel blocks $s_\lam$, $0 \leq \lam \leq m_{kI}$)
coincide for $I = \frac 32$ and $\frac 72$. It is instructive
to verify that, although the $s$-channel $F$-matrices do not
coincide for $I = \frac 72$ and $\frac 32$, the invariant ratios
(3.15) are the same.

In computing $F\sch = US\inv$ for $I=\frac 72$ in terms of the $s$- and
$u$-channel transition matrices $S$ and $U$ (see Sect.\ 2), one
encounters the problem of the reduction from the 8-dimensional space of
KZ solutions to the 4-dimensional physical subspace. It is simplified by
the observation that due to the triangular form of $S$ and $U$, the
reduced matrix $F\sch$ for $4I>k$ is obtained by just taking the first
$m_{kI}+1 = k-2I+1$ rows and columns of both $U$ and $S\inv$. In
particular, for $I = \frac 72$ we observe that the symmetrized (unitary)
matrices (3.10) corresponding to $I = \frac 32$ (eq.\ (3.13)) and to $I
= \frac 72$ coincide.

The 2-dimensional braid invariant subspace comprising the conformal
blocks of local Ramond fields of the $\bb_{10}$ model is spanned by the
pair of vectors
\be v_0 = \textstyle (-\frac23,0,0,\sqrt\frac23 ), \quad v_2 =
(0,0,1,0) \ee
which are ortho-normalized with respect to the metric (3.14):
\be {^tv}_a\DD v_b = \d_{ab} \qquad (a,b = 0,2). \ee
In this basis, we have the following reduced form of the $s$-channel
generators
\be \hat B_1\sch = q^{\frac 32} \pmatrix{q^3&0\cr 0&-q^{-3}\cr},
\qquad \hat F\sch = -\frac {[2]}{[3]}\pmatrix{-1&1\cr 1&1\cr}
\qquad\quad (k=10, I = \textstyle\frac32). \ee
(At $q = e^{\frac{i\pi}{12}}$, one has $[3] = \sqrt2[2]$).
Identical expressions are obtained for the reduced generators acting
in the invariant subspace of \cfb s for $I = \frac 72$.

The resulting 2-dimensional \rep\ of $\BBB_4$ is a finite matrix group.
It is a central extension of the 24-element 2-fold covering of the
tetrahedron group. This is worth noticing, since the appearance of
finite matrix groups among the monodromy \rep s of $\BBB_4$ is rather
exceptional \cite{ST}.
\sectreset{Subfactors for field extensions}
We turn now to the treatment of the same problem: the determination
of relative amplitudes like (3.8), (3.9), in the algebraic (DHR)
framework of \qft. A theory $\aa$ is described by a local net of
\vna s $\aa(\oo)$ of observables in the space-time region $\oo$, which
generate the global $C^*$ algebra $\aa$. These regions may be double
cones ($\oo$), or intervals ($\ii$) on the \lc\ in chiral conformal
theories.

In the following, we consider a pair of local quantum field theories
given by the nets of local \vna s $\aa(\oo)$ and $\bb(\oo)$ such that
\be \aa(\oo) \subset \bb(\oo) \ee
are irreducible inclusions with common unit. Our terminology will be
`observables' for $a \in \aa$ and `charged fields' for $b \in \bb$.
We have in mind two specific such nets, namely
\begin{enumerate} \item the conformal inclusion \cite{FGPFKS,MST} of
the chiral $su(2)$ \ca\ at level 10 into the chiral $sp(4)$ \ca\ at
level 1, denoted by
\be \aa\ch(\ii) \subset \bb\ch(\ii) \ee
where $\ii$ are intervals on the circle (= compactified conformal
\lc), and
\item the two-dimensional WZNW model \cite{WZW} of the chiral $su(2)$
currents at level 10 (on both \lc s) contained in the algebra of
two-dimensional local fields constructed by diagonal contraction of
chiral vertex operators (exchange fields):
\be \aa\two(\oo) \equiv \aa\ch(\ii) \otimes \aa\ch(\bar \ii)
\; \subset \; \ff\two(\oo) \ee
where a two-dimensional double cone $\oo = \ii \times \bar \ii$
is the Cartesian product of two chiral \lc\ intervals.
\end{enumerate}
Note that the model (4.3) is the one described by the standard diagonal
form $D$ in the previous sections, while the form $\DD$ corresponds
to a combination of (4.2) and (4.3):
\[ \aa\two(\oo) \equiv \aa\ch(\ii) \otimes \aa\ch(\bar \ii)
\; \subset \; \bb\ch(\ii) \otimes \bb\ch(\bar \ii) \; \subset \;
\wt \ff\two(\oo). \]
Here, the first inclusion is the tensor product of the chiral extensions
(4.2) and the second inclusion is the standard diagonal contraction of
chiral vertex operators for $\bb\ch$. (There will be said more about
these `standard' constructions in Sect.\ 5; see also \cite{Spt,FLRR}.)

A subfactor $\sub AB$ is irreducible if the relative commutant is
trivial: $A' \cap B = \CC$. This requirement excludes from our analysis
all chiral current subalgebras associated with subgroups, unless the
embedding is `conformal' \cite{BNSW}, since the coset \emt\ is contained
in the relative commutant. However, including the coset \emt\ into the
observables (which then have the structure of a tensor product of two
chiral theories), would again yield an irreducible inclusion
\cite{Wass,FLRR}.

We have to recall some subfactor theory. First, we note that we are
dealing with type $\III_1$ subfactors, since under very general
conditions, the local \vna s in \qft\ are \hpf\ type $\III_1$ factors
\cite{BDF,FG}. Associated with an (irreducible) type $\III$ subfactor
$\sub AB$ is a {\it \cmor} $\g \in End(B)$ such that $\g(B) \subset A$
is a dual subfactor \cite{Lo,Ko}. $\sub AB$ has finite index if and
only if \cite{Lo} there is a pair of isometries $W \in A$ and $V \in B$
such that the following operator identities hold:
\be \barr{ll} (a) & Wa = \rho(a)W \qquad\quad
(a \in A,\,\rho:=\g\vert_A) \\
(b) & Vb = \g(b)V \quad\qquad\qquad (b \in B) \\
(c) & W^*V = \lam^{-1/2} \eins = W^*\g(V). \earr \ee
The real number $\lam$ is called the {\it index} of the subfactor $\sub
AB$. These relations express the duality between $\sub AB$ and $\g(B)
\subset A$. They also state that $B$ is the Jones extension \cite{Jo} of
$A$ by its subfactor $\g(B)$. The Jones projection is $E=VV^*$,
satisfying the Jones-Temperley-Lieb relation with its dual $F=WW^*$:
\[ EFE = \lam\inv E, \qquad FEF = \lam\inv F. \]
Associated with these data, there is a {\it \cex} $\mu \map B A$ given
by
\be \mu(b) = W^*\g(b)W \qquad (b \in B), \ee
and conversely the \cmor\ can be expressed in the form
\be \g(b) = \lam\cdot\mu(VbV^*) \qquad (b \in B) . \ee
$\mu$ is a positive and $A$-linear map which generalizes the Haar
average over a compact group acting on $B$ with fixpoints $A$. It
satisfies the \pipo\
\be \mu(b) \geq \lam\inv \cdot b  \qquad (b \in B,\;  b \geq 0) \ee
as an operator estimate for every positive operator $b \in B$. This
lower bound for \cex s was first introduced in \cite{PiPo} to define
the index. It is optimal since it is saturated by
\[ \mu(VV^*) = \lam\inv \eins. \]
We note also that $W = \lam^{-1/2} \cdot \mu(V)$. The physical relevance
of these objects will become clear in due context.

The following results on quantum field theoretical nets of subfactors
as in eq.\ (4.1) will be proven (and qualified) elsewhere \cite{FLRR}.
Let us just state the essentials. Let the vacuum vector $\Omega$ be
cyclic and separating for every local \vna\ $\bb(\oo)$ of the theory
$\bb$, i.e.\ $\pi^0(\bb(\oo))\Omega$ are dense subspaces of the \vrep\
space $\hh^0$. This property holds, by the Reeh-Schlieder Theorem, quite
generally for covariant quantum field theories with positive energy.
Let also $\hh_0 = \ovl{\pi^0(\aa)\Omega} \subset \hh^0$ be the \vrep\
space of $\aa$ such that $\Omega$ is also cyclic and separating in
$\hh_0$ for every $\aa(\oo)$. Let furthermore the \cex\ $\mu$ preserve
localizations, i.e., map $\bb(\oo)$ onto $\aa(\oo)$. If the local
subfactors are irreducible and therefore possess a unique \cex, then
$\mu$ must commute with the translations (= the rotations of the
circle in the case of a chiral conformal theory). If the vacuum state
$\omega = \LO,\pi^0(\cdot)\RO$ on $\bb$ is the unique translation
invariant state, then it must also be invariant under $\mu$, i.e.,
\be \omega\comp\mu = \omega \qquad {\rm on} \quad \bb. \ee
We shall assume the invariance property (4.8) in the sequel. The
underlying structure admits the interpretation as a generalized global
unbroken gauge symmetry with $\mu$ generalizing the gauge group average
\cite{FLRR}.

Under these circumstances, the \cmor\ $\g$ defined above for a fixed
local subfactor $\aa(\oo_0) \subset \bb(\oo_0)$ extends to an \emor\ of
the global $C^*$ algebra $\bb$, and maps $\bb$ into the global $C^*$
algebra of observables $\aa$. Restricted to the observables,
$\g\vert_\aa$ turns out to be a localized \emor\ with localization in
$\oo_0$, denoted by $\rho$ in the sequel. It therefore describes a
(reducible) \sps\ sector \cite{DHR} of the theory $\aa$. Its physical
significance is given by the following
\bq
{\bf Proposition 4.1:} \cite{Wass,FLRR} Let $\po$ denote the \vrep\ of
$\aa$ on $\hh_0$, and $\pi^0$ the \vrep\ of $\bb$ on $\hh^0$. Then
$\pi^0$ considered as a reducible \rep\ of the subalgebra $\aa$ is
unitarily equivalent to the \rep\ $\po\comp\rho$ of $\aa$.
\eq
In other words: the \sps\ sector $\rho$ comprises all the charged
sectors of $\aa$ which are interpolated from the vacuum by fields in
$\bb$. If, as \emor s, $\rho \simeq \bigoplus_s N_s\rho_s$, then as
\rep s,
\be \pi^0\vert_\aa \simeq \po\comp\rho \simeq \bigoplus_s N_s\pi_s \ee
where $N_s$ are finite multiplicities, and $\pi_s \equiv
\po\comp\rho_s$. As is well known, if the observables $\aa$ are the
gauge invariants under a compact gauge symmetry group of $\bb$, then
the decomposition (4.9) is given by the \rep s of the gauge group,
with multiplicities $N_s$ given by the dimensions of the latter.

Eq.\ (4.9) allows to compute the index $\lam$ of the subfactor.
It is given by the formula
\be \lam = d(\rho) = \sum_s N_s d(\rho_s) \ee
in terms of the statistical dimensions $d(\rho_s) \equiv d_s$
of the \sps\ sectors \cite{DHR} contained in $\rho$. In the gauge group
case, $d(\rho_s) = N_s$, and the index equals the order of the group.

In the models (4.2), (4.3), the branching of the vacuum sector of $\bb$
is well known, leading to $\rho \simeq \rho_0 \oplus \rho_3$ for the
inclusion (4.2) and $\rho \simeq \bigoplus_I \rho_I \otimes \rho_I$ for
the inclusion (4.3), where $\rho_I$ are the isopin $I$ sectors of the
chiral $su(2)$ \ca. $\rho_0 \equiv \id$ corresponds to the \vrep. In
the former case, the formula (4.10) yields the index $\lam = d_0 + d_3
= 1+\sin\frac{7\pi}{12}/\sin\frac\pi{12} = 3+\sqrt3$. (For the
coincidence of statistical dimensions and `quantum dimensions'
$d(\rho_I) = [2I+1]$ for $su(2)$ \ca s see \cite{Wass}.)

The formulae (4.4) -- (4.7) remain valid for $\g$ considered as an
\emor\ of $\bb$ and for $\rho$ as an \emor\ of $\aa$. Note that the
isometries $W$ and $V$ are local operators $W \in \aa(\oo_0)$ and $V
\in \bb(\oo_0)$. We shall refer to the intertwining properties expressed
by eqs.\ (4.4($a,b$)) by the notation $V \map \id \g$ and $W \map \id
\rho$ in the sequel. The latter implies that $\po(WW^*)$ is the
projection in the \rep\ space of $\po\comp\rho$ which projects onto the
vacuum sub\rep\ contained in (4.9).

For every other subsector $\pi_s$ contained in (4.9) there are
corresponding projections of the form $\po(W_{s,i}W_{s,i}^*)$ where
$W_{s,i} \map {\rho_s}\rho$ are orthonormal isometric \itw s in
$\aa(\oo_0)$; the multiplicity index $i$ runs from 1 to $N_s$. For
simplicity, we shall in the following consider only multiplicities $N_s
= 1$ (covering abelian gauge groups, as well as our models above). One
has the orthogonality relation $W_s^*W_t = \d_{st}$ (because otherwise,
the \itw\ $W_s^*W_t \map {\rho_t}{\rho_s}$ would contradict the
inequivalence of the \rep s $\pi_s$ and $\pi_t$), and the completeness
relation $\sum_s W_sW_s^* = \eins$. Clearly, $W_0 \equiv W$.

Putting
\[ \psi_s := W_s^*V  \]
we obtain charged \itw s, i.e., elements of $\bb$ which satisfy the
\comrel s with the observables
\be \psi_s a = \rho_s(a) \psi_s \qquad (a \in \aa) .\ee
This equation means that $\psi_s \in \bb$ make transitions (in the
\vrep\ of $\bb$) between the \vrep\ of $\aa$ and the charged \rep s
$\pi_s$.

Conversely,
\be V = \sum_s W_s\psi_s , \ee
and the \comrel\
\be Va = \rho(a)V \qquad (a \in \aa) \ee
gives to $V$ the physical interpretation as a `master field' carrying
the reducible charge $\rho$ from which the charged \itw s $\psi_s$ are
projected out by means of $W_s$.

A particularly interesting object is the observable operator
\be X := \g(V) \in \aa(\oo_0) . \ee
{}From the definitions it is clear that $X$ is an isometric \itw\
$X \map \rho {\rho^2}$. Indeed, we can compute
\[ X = \g(V) = \lam\mu(VVV^*) = \lam\sum_{stu}\mu(W_t\psi_t
W_s\psi_s\psi_u^*W_u^*) = \lam\sum_{stu}W_t\rho_t(W_s) \cdot
\mu(\psi_t\psi_s\psi_u^*) \cdot W_u^* \]
where the expressions $\mu(\psi_t\psi_s\psi_u^*)$ are observable \itw s
$T \map {\rho_u}{\rho_t\rho_s}$. They are therefore multiples of
isometric basis \itw s $T_e$ which project onto the sub\rep s $\pi_u$
contained in the DHR composition product $\pi_t \times \pi_s =
\po\comp(\rho_t\rho_s)$:
\be \lam\mu(\psi_t\psi_s\psi_u^*) = \lam(e) \cdot T_e \ee
with coefficients
\be \lam(e)\; \eins = \lam \cdot T_e^*\mu(\psi_t\psi_s\psi_u^*). \ee
(The multi-index $e$ stands here and in the sequel for the fusion
channel $\pi_u \prec \pi_t\times\pi_s$.)

Denoting by $\tilde T_e = \rho(W_s) W_t\cdot T_e \cdot W_u^*$ the
`lifts' of \itw s $T_e \map {\rho_u}{\rho_t\rho_s}$ to \itw s
$\tilde T_e \map \rho\rho^2$, we obtain the expansion
\be X = \sum_e \lam(e) \; \tilde T_e. \ee

We note that only channels $e$ contribute to (4.17) for which
$\rho_s,\rho_t,\rho_u$ are all subsectors of the \cmor\ $\rho$, in
spite of the fact that in general $\rho_t\rho_s$ will also contain
subsectors which are not contained in $\rho$. We shall relate this
observation to the `truncated fusion rules' in the next section.

The importance of the isometry $X$ is due to the following result,
while the relevance of its expansion coefficients $\lam(e)$ will
reveal itself in the sequel.
\bq
{\bf Proposition 4.2:} \cite{Hopf} The irreducible subfactor $A \subset
B$ is uniquely characterized (up to unitary equivalence) by the triple
$(\rho,W,X)$, where $\rho \in End(A)$ and $W \map \id\rho$ and
$X \map \rho\rho^2$ are isometric \itw s in $A$, satisfies the
following identities
\be \barr{lcl} (i) & W^*X = \lam^{-1/2}\eins = \rho(W^*)X & \quad
{\rm with} \quad \lam = d(\rho) \\ (ii) & XX^* = \rho(X^*)X & \\
(iii) & XX = \rho(X)X. & \earr \ee
\eq
Clearly, the identities (4.18) follow from (4.4). Conversely, given a
triple as in Prop.\ 4.2, one recovers $B$ as follows. Put $A_1 :=
X^*\rho(A)X$ and $B :=$ the Jones extension of $A$ by $A_1$. This
extension is of the form $B = AV$ where $V$ is an isometry with $VV^* =
E$, the Jones projection. Define $\g \in End(B)$ by $\g(aV):=\rho(a)X$.
Then $\g$, satisfying (4.4), is the \cmor\ for $A \subset B$ and $\rho
= \g\vert_A$, $A_1 = \g(B)$.

In our present context, $A = \aa(\oo)$ and $B = \bb(\oo)$, the point
about this characterization of (4.1) is that it entirely refers to
the observables and their \sps\ sectors. Finding such a triple in a
given theory $\aa$ amounts to find a field extension $\bb$ of the
observables of the form (4.1). The problem involves the knowledge of the
`fusion coefficients' of the theory $\aa$, i.e., the coefficients of
expressions like $\rho_v(T_e)$ (entering $\rho(X)$) in terms of a basis
$T_gT_hT_f^*$. These are the solutions to the Moore-Seiberg `pentagon
identities' \cite{MS} which are intrinsically determined by the DHR
theory of \sps\ sectors \cite{FRSII} (but often tedious to compute).

Let us briefly sketch the `reverse program' of construction and
classification of (local) field extensions of finite index \cite{FLRR}.

The main step is to decide which combinations $\rho \simeq \bigoplus_s
N_s \rho_s$ of the irreducible localized \emor s (sectors) of $\aa$ are
{\it canonical} \emor s of the local \vna\ $A \equiv \aa(\oo_0)$ with
respect to some subfactor $A_1 \subset A$. This amounts \cite{Hopf} to
verify the existence of a pair of isometric \itw s $W \map \id\rho$ and
$X \map \rho\rho^2$ in $\aa(\oo_0)$ solving (4.18). If the desired
inclusion is required to be irreducible, then $id \prec \rho$ with
multiplicity $N_0 = 1$, and if the index is finite, then one can prove
the bound $N_s \leq d_s$. Therefore, if $\aa$ is a `rational' theory,
i.e., has only finitely many sectors of finite statistics, then the
classification problem is a finite problem in the form of a non-linear
system for the unknown coefficients $\lam(e)^k_{ij}$ (with
multiplicities).

If we are interested in {\it local} field extensions, then we have to
require in addition (see below) that the solution $X$ satisfies
\be \eps_\rho X = X \ee
where $\eps_\rho \in \rho^2(\aa)' \cap \aa(\oo_0)$ is the \st\ operator
for the localized \emor\ $\rho$ \cite{DHR}. $\eps_\rho = U^*\rho(U)$
can be computed in terms of a charge transporting intertwiner $U \map
\rho \hat\rho$ where $\hat\rho$ is an equivalent \emor\ localized at
space-like distance from $\rho$.

Every solution $(\rho,W,X)$ to the system (4.18) defines a field net
$\bb$ extending $\aa$ with finite index $\lam = d(\rho)$ as follows. If
$\rho$ is localized in $\oo_0$, one reconstructs $B = \bb(\oo_0)$
and $\g \in End(B)$ from $A = \aa(\oo_0)$ as in the remark after
Prop.\ 4.2. Thus $\bb(\oo_0) = \aa(\oo_0)V$ for an isometry $V \in
\bb(\oo_0)$ satisfying (4.4). Next, $\bb(\oo) := \aa(\oo) UV$ are
defined with the help of charge transporters $U \in \aa$, i.e., unitary
\itw s $U \map \rho\hat\rho$ where $\hat\rho$ is localized in $\oo$.
Note that $\bb(\oo)$ thus defined contains the identity operator $\eins
\propto W^*V = \hat W^*UV$ since $\hat W = UW \map \id\hat\rho$ is in
$\aa(\oo)$. Consequently, $\bb(\oo)$ contains and extends $\aa(\oo)$.
This construction yields a net $\bb$ which is relatively local with
respect to $\aa$, since $\rho$ is localized; namely if $\oo$ is at
space-like distance from $\oo_0$, then $\aa(\oo_0)$ commutes with
$\bb(\oo)$:
\[ UV \cdot a = U \rho(a) V = \hat\rho(a) UV = a \cdot UV \qquad
(a \in \aa(\oo_0)). \]
The field extension $\bb$ turns out to be local if and only if the
solution $X$ satisfies also (4.19). Namely, the commutativity of
$V \in \bb(\oo_0)$ with $UV \in \bb(\oo)$ at space-like distance
is equivalent to $VV = U^*VUV$, and hence to
\[ XV = \g(V)V = VV = U^*VUV = U^*\rho(U)VV =
\eps_\rho\g(V)V = \eps_\rho XV. \]

We observe that the system (4.18) alone will have many solutions,
e.g., those of the form $\rho = \bsig\sig$, $X = \sig(\bar W)$
where $\sig$ is any irreducible localized \emor\ of the theory $\aa$
with finite \st s, $\bar W \map \id\sig\bsig$ an isometry. These
solutions will, however, not satisfy (4.19) in general, and will
therefore not give rise to local field extensions.

Note that, actually, locality of the field net was not required for the
general analysis in the first part of this section, as long as it has
the Reeh-Schlieder property, and fields commute with observables at
space-like distance. However, since it is not clear which physical
principles should determine a `good choice' of a non-local and
therefore {\it a priori} unobservable field algebra except that it
generates the \sps\ sectors of the observables, we prefer to consider
only local field extensions which offer the option to be regarded as
observable theories of their own.

If $\aa$ are the gauge invariants under a gauge group acting on $\bb$,
then the system (4.18) has a solution with multiplicities $N_s$ given by
the dimensions of the \rep s of the gauge group. The corresponding
coefficients $\lam(e)^k_{ij}$ in the expansion (4.17) of $X$ are
precisely the \CGC s. Indeed, one may rephrase the content of the
Doplicher-Roberts (DR) reconstruction theorem \cite{DR}
as follows: every system of sectors of the observables which have finite
permutation group statistics among each other, closed under composition,
reduction, and conjugation, admits a solution to (4.18) with $X$ given
by (4.17) in terms of \CGC s of some compact gauge group. The DR
solution is distinguished by the validity of (4.19) if
there are only bosonic sectors of $\aa$, and a graded variant of (4.19)
in the presence of fermionic sectors.

We emphasize that, while our general theory above contains the case
of a compact gauge symmetry group, the models (4.2), (4.3) we are
actually interested in are not given by a gauge symmetry group. The
sectors $\pi_s$ contained in the restriction $\pi^0\vert_\aa$ are not
closed under composition, and their multiplicities differ from their
statistical dimensions. Although the fields are local, the sectors
$\pi_s$ have braid group statistics. None of these features could
hold with a gauge group.

Displayed in terms of the coefficients $\lam(e)^k_{ij}$, the system
(4.18) is converted into a system of identities well-known to hold for
\CGC s (with the $6j$ symbols as fusion coefficients). The absence of a
completeness property in (4.18) is
related to the truncated fusion rules discussed in the next section.

\sectreset{Truncated fusion rules and partial wave decomposition}
Let us now study multiplicative properties of the charged
fields $\psi_s$ (`operator product expansions'). For a generic
charged operator $b \in \bb$ one has the expansion formula
(generalizing the harmonic analysis in the gauge symmetry case)
implied by (4.4), (4.5)
\be b = \lam\mu(bV^*)V = \lam\sum_s\mu(b\psi_s^*)\psi_s \qquad
(b \in \bb). \ee
In particular, by (4.15),
\be \psi_t\psi_s = \sum_u \lam(e)\; T_e\psi_u \ee
where as before, $e$ is the channel $\rho_u \prec \rho_t\rho_s$. We
observe, that only charged fields with charge $\rho_u \prec \rho$
contribute to this operator product expansion, even if there are other
sectors present in the DHR sector decomposition of $\rho_t\rho_s$.
That this `truncation of the fusion rules' is consistent, can be
retraced, e.g., to the identity (4.18($iii$)) as follows.

Obviously, $\psi_t\psi_s$ is a charged \itw\ $\map \id\rho_t\rho_s$,
so one might expect that all charges $\rho_v$ contained in
$\rho_t\rho_s$ are interpolated by this composite field. But, in order
to project a field carrying charge $\rho_v$ out of $\psi_t\psi_s$, we
have to multiply the latter with $T_e^*$ where $T_e \map {\rho_v}
{\rho_t\rho_s}$. Now, computing $T_e^*\psi_t\psi_s$, or rather its image
under $\rho$, we get
\[ \rho(T_e^*\psi_t\psi_s) = \rho(T_e^*W_t^*VW_s^*V)
= \rho(T_e^*W_t^*\rho(W_s^*) \cdot VV) =
\rho(T_e^*W_t^*\rho(W_s^*))\cdot XX \]
Using $XX = \rho(X)X$, we obtain an expression involving
$\rho[T_e^*W_t^*\rho(W_s^*)X]$ where the argument in square brackets
is an \itw\ $\map \rho\rho_v$ in $\aa$ which must vanish unless
$\rho_v\prec\rho$. In other words, since the expansion (4.17) of
$X$ contains only $T_e$ for fusion channels which are already
contained in $\rho$, it is annihilated by all $T_e$ leading to
other channels. Therefore, the identity $T_e^*\psi_t\psi_s = 0$
following from identity ($iii$) precisely describes in the operator
product expansion for charged fields the suppression of channels
$\rho_v$ not contained in $\rho$, i.e., the truncated fusion rules.

We now turn to our main result, the decomposition of correlation
functions of charged fields into `partial wave' contributions, and the
decomposition of charged fields $\psi_s$ into `chiral exchange fields'.

Applying the expansion (5.2) (and (4.11)) repeatedly, we find
the following expansion for vacuum correlations of generic charged
fields of the form $\varphi = \psi_s^*a$
\be \LO, \vphi_n \cdots \vphi_1 \RO = \sum_\xi \prod_i \ovl{\lam(e_i)}
\cdot \LO, T_{e_n}^*\rho_{t_n}(a_n) \cdots T_{e_2}^*\rho_{t_2}(a_2)
T_{e_1}^*a_1 \RO \ee
where $T_{e_i} \map {\rho_{u_i}}{\rho_{t_i}\rho_{s_i}}$ and the sum
extends over all vacuum-to-vacuum `channels' of successive fusion $\xi
= e_n \comp \cdots \comp e_1$ such that $t_i = u_{i-1}$ and $u_n = 0 =
t_1$. The last step in this computation, the evaluation of a single
charged field of the form $\psi_s^*a$ in the vacuum state, exploits
the invariance of the vacuum state
\[ \omega(\psi_s^*a) = \omega(\mu(\psi_s)^*a) = \d_{s0}\lam^{-1/2}
\omega(a) \]
since $\mu(\psi_s) = W^*\g(W_s^*V)W = W_s^*W^*XW = \d_{s0}\lam^{-1/2}
\eins$. The factor $\lam^{-1/2}$ is absorbed in the product in
(5.3) in the guise of $\lam(e_1)$ (note that for $\rho_t = \id$, $T_e
= \eins$, and $\tilde T_e = W W_sW_s^*$, one obtains $\lam(e) =
W_s^*W^*XW_s = \lam^{-1/2}$).

In the formula (5.3), the single `partial wave' contributions
\be F_\xi = \LO, T_{e_n}^*\rho_{t_n}(a_n) \cdots T_{e_2}^*
\rho_{t_2}(a_2) T_{e_1}^*a_1 \RO \ee
are kinematically distinguished correlation functions which depend only
on the subtheory $\aa$ and its \sps\ structure, but bear no reference to
the field extension $\bb$.
\bq
{\bf Proposition 5.1:} The (local) $n$-point functions of charged fields
from a field extension $\bb$ have the partial wave expansions (5.3)
where only the coefficients
\be N_\xi = \prod_i \ovl{\lam(e_i)}, \ee
involving the factors $\lam(e_i)$ for every single transition in the
channel of successive fusions, depend on $\bb$.
\eq
On the Hilbert space of the \rep\ $\pi^0\vert_\aa \equiv \bigoplus_s
\pi_s$ (cf.\ Prop.\ 4.1) were also defined the `\rfb' operators $F(e,a)$
as a bounded operator version of chiral vertex operators
\cite{FRSI,FRSII}. If $e$ is the channel $\rho_u \prec \rho_t\rho_s$,
then $F(e,a) \equiv F(e,\eins)\pi^0(a)$ interpolates from the subspace
for $\pi_t$ to the subspace for $\pi_u$ by the formula
\[ F(e,a) |t;\Psi\rangle := |u;\po(T_e^*\rho_t(a))\Psi\rangle. \]
These operators satisfy complicated `exchange algebra' \comrel s
(whence the name `exchange fields' \cite{EA,Spt}) involving matrix
elements of the relevant \st\ operators (braid matrices), and a
multiplication law involving the fusion coefficients for the sectors.
The algebra spanned by $F(e,a)$ is closed under multiplication and under
the adjoint operation.

By inspection of the partial wave contributions (5.4) one sees that
the latter are just the correlation functions of products of \rfb\
operators $F(e,a)$. Therefore (5.3) implies the identification
\be \psi_s^*a = \sum_e \ovl{\lam(e)}\; F(e,a) \ee
where the sum extends over all fusion channels with fixed charge label
$s$. This formula is remarkable since the charged fields in $\bb$ which
satisfy local \comrel s and truncated fusion rules as discussed above,
arise as specific linear combinations of \rfb\ operators which satisfy
exchange algebra \comrel s and do not exhibit truncation. Similarly,
while every single partial wave contribution (5.4) is non-local, the
sum (5.3) is a local $n$-point function. This is possible due to
cancellations among the relevant fusion coefficients, which can be seen
to follow from the system (4.18), (4.19) if written as a nonlinear
system involving fusion coefficients and braid matrices along with the
\CGC s $\lam(e)$. A similar statement applies to the identities
\[ \psi_s^* = d_s^{1/2}\; R_s^* \psi_{\bar s} \qquad
(R_s \map \id\brho_s\rho_s \; {\rm isometric}) \]
and
\[ \psi_s^*\psi_s = d_s/\lam \cdot \eins \]
valid in $\bb$, which we have not discussed here, but which can be
proven within the \rfb, with the identification (5.6), along the same
lines. We refrain from working out the details here.

Actually, the decomposition (5.6) can also be directly established in
terms of the unitary equivalence between $\bigoplus \pi_s$ and
$\pi^0\vert_\aa$.

It was argued in \cite{FRSII} that in a sufficiently regular theory
with conformal covariance, scaling limits of $F(e,a)$ contracting the
localization to a point $x$ should exist, and yield chiral vertex
operators $\vphi_e(x)$ interpolating between the sectors $\hh_t$ and
$\hh_u$:
\[ \vphi_e(x) \sim \lim_{\lam\arr 0} \lam^{-\Delta_s}
F(e,\a\sch_x\a\sch_\lam(a)) \]
where $\a\sch$ denotes the charge dependent effect of the translation
($x$) resp.\ scale transformation ($\lam$) on the operator entry $a$
\cite{FRSII}. It follows from these considerations that the pointlike
limits of $\psi_s^*a$ yields local pointlike fields affiliated with
$\bb$ of the form
\be \psi_s(x) = \sum_e \ovl{\lam(e)}\; \vphi_e(x) \ee
with the same coefficients as in (5.6). E.g., in the model (4.2) the
heptuplett of primary currents $j^a(x)$ for the isospin 3 sector arise
as linear combinations of vertex operators with coefficients $\lam(e)$
to be computed below, and the same holds in general for charged local
fields from $\bb$.

In the pointlike limit, the partial wave contributions tend to
`\cfb\ functions'
\be \ff_\xi(x_n,\ldots,x_1) = \LO,\vphi_{e_n}(x_n) \cdots
\vphi_{e_1}(x_1)\RO \ee
to be identified with the $s$-channel blocks in the standard approach,
determined from Ward identities and Knizhnik-Zamolodchikov equations
\cite{KZ}. For $n=4$, the label $\xi$ stands just for the intermediate
sector due to $s$-channel fusion of the charges $s_1$ and $s_2$.
Although the \cfb s are non-local functions, their combinations with
coefficients as in (5.3) are local $n$-point functions of local fields
like $j^a(x)$.

The limit behaviour of partial waves (5.4) tending to \cfb s
(5.8) is an intrinsic property of the subtheory $\aa$. As in Prop.\
5.1, $n$-point functions of local charged fields $\psi_s(x)$ depend on
the field extension $\bb$ only through the expansion coefficients
$N_\xi$ given by (5.5).

In practice, we don't know the absolute normalizations of the limiting
functions (5.8) in order to identify them with a choice of $s$-channel
solutions as in Sects.\ 2 and 3, nor do we know the relative
normalizations of different partial wave contributions with respect to
each other. This would require the control of the previously mentioned
pointlike limits which is in general a difficult problem. However, one
can compute double ratios, which compare two different field
extensions, of the form
\be \frac{N'_\xi/N'_\eta}{N_\xi/N_\eta} \ee
which are completely normalization independent `characteristic'
quantities. These double ratios must in particular coincide with the
corresponding double ratios comparing relative amplitudes of
$s$-channel \cfb\ functions contributing to $n$-point
functions of point-like fields from two different field extensions, as
computed in Sects.\ 2 and 3.

Since the double ratios (5.9) are given by (5.5), we have established
the desired relation between relative amplitudes of \cfb s and
the data of the relevant local subfactors. This relation is based on
the identification of the expansion coefficients in (4.17) for the
characteristic isometry and in (5.2) for operator products of charged
fields (reflected also in (5.6) for charged fields as elements of
the \rfb).

Let us now compute the amplitudes (5.5) for our first model (4.2) from
the characteristic triple $(\rho,W,X)$. From the branching of the \vrep\
of $\bb$ upon restriction to $\aa$, we know that $\rho \simeq \rho_0
\oplus \rho_3$ (see Sect.\ 4). By (4.10), the index is $\lam = d(\rho)
= d_0+d_3 = 3+\sqrt3$. Actually, finite index type $\III_1$ subfactors
are isomorphic to type $\II_1$ subfactors tensored with a type $\III$
factor \cite{Popa}. The corresponding type $\II_1$ subfactor associated
with the model (4.2) is the well known subfactor of index $\lam =
3+\sqrt3$ constructed in \cite{GHJ}.

Choosing $\rho_0 = \id$ in its equivalence class, the isometry $W \map
\id \rho$ is uniquely determined up to an irrelevant phase. The
coefficients $\lam(e)$ for the isometry $X$ can be computed from $X^*X
= \eins$ and the identity (4.18($i$)): there are only five fusion
channels $\rho_u \prec \rho_t\rho_s$ with all $\rho_s,\rho_t,\rho_u
\prec \rho$, with which we associate isometric \itw s as follows:
\[ T_a \map {\rho_0}{\rho_0\rho_0}
                       \qquad T_b \map {\rho_3}{\rho_3\rho_0}
\qquad T_c \map {\rho_3}{\rho_0\rho_3}
                       \qquad T_d \map {\rho_0}{\rho_3\rho_3}
\qquad T_e \map {\rho_3}{\rho_3\rho_3} \]
Since $\rho_0 = \id$, we may choose $T_a = T_b = T_c = \eins$.
According to standard notation \cite{DHR,FRSI,FRSII}, we call $R$ the
isometry $T_d \map \id\rho_3^2$. We have therefore:
\[ \barr{l} X = \lam(a)\cdot\rho(W_0)W_0W_0^* + \lam(b)\cdot
\rho(W_0)W_3W_3^* + \\ \quad + \lam(c)\cdot\rho(W_3)W_0W_3^*
+ \lam(d)\cdot\rho(W_3)W_3RW_0^* + \lam(e)\cdot\rho(W_3)W_3 T_e W_3^*,
\earr \]
where $W_0 \equiv W \map \id\rho$ and $W_3 \map{\rho_3}\rho$ are
orthonormal isometries, and $E_0 = W_0W_0^*$ and $E_3 = W_3W_3^*$
are complementary projections in the commutant of $\rho$ onto the two
subsectors of $\rho$. Then (4.18($i$)) reads
\[ W_0^*X = \lam(a)E_0 + \lam(c)E_3 = \lam^{-1/2} \eins \]
\[ \rho(W_0^*)X = \lam(a)E_0 + \lam(b)E_3 = \lam^{-1/2}
\eins \]
hence $\lam(a) = \lam(b) = \lam(c) = \lam^{-1/2}$. We are free to
choose the complex phases of $R$ and $T_e$ such that $\lam(d)$ and
$\lam(e)$ are also positive. Now, the isometricity of $X$ together with
the orthogonality of $R$ and $T_e$ (i.e., $R^*T_e = 0$) implies
\[X^*X = [\lam(a)^2+\lam(d)^2]E_0 +
[\lam(b)^2+\lam(c)^2+\lam(e)^2]E_3  = \eins \]
hence $\lam(d) = \sqrt{1-\lam\inv}$ and $\lam(e) =
\sqrt{1-2\lam\inv}$.
We don't need to verify the remaining identities (4.18), (4.19) since we
know that the extension is local and yields a subfactor of index $\lam =
3+\sqrt 3$. (Unfortunately, the computation is much less trivial for the
other, $E_8$, extension treated in Sects.\ 2 and 3.)

For charged fields with $\rho_s = \rho_3$, only the channels $c \equiv
(30)$, $d \equiv (03)$, $e \equiv (33)$ are relevant ($(JI)$ stands for
an exchange field of charge 3 acting on $\hh_I$ with values in
$\hh_J$.) Therefore, we have
\[ \lam(30) = \lam^{-1/2},\qquad \lam(03) = \left(\frac{\lam-1}\lam
\right)^{1/2}, \qquad \lam(33) = \left(\frac{\lam-2}\lam\right)^{1/2}.
\]
This gives for the ratio of the amplitudes of the \cfb s
with intermediate $s$-channel $I=0,3$ contributing to the
4-point function of the isospin 3 field
\be N_3/N_0 = \frac{\lam(03)\lam(33)\lam(33)\lam(30)}
{\lam(03)\lam(30)\lam(03)\lam(30)} = \frac{\lam-2}{\sqrt{\lam-1}} =
\sqrt2. \ee

As discussed before, due to uncontrolled normalizations, one has to
compute double ratios like (5.9) of relative amplitudes comparing two
different field extensions. Indeed, there is always a `standard'
extension to compare with, which specializes for chiral \ca s to the
$A$ series of modular invariants \cite{CIZ}, and therefore yield the
diagonal extensions as in our model (4.3).
\bq
{\bf Proposition 5.2:} \cite{Plek,FLRR} For rational chiral theories
$\aa\ch$ (i.e., theories with only a finite number of \sps\ sectors
$\pi_s$ with finite statistics), $\rho \simeq \bigoplus_s \rho_s
\otimes \brho_s$ is a \cmor\ of $\aa\two \equiv \aa\ch \otimes \aa\ch$,
giving rise to a local two-dimensional field extension $\bb\two$.
\eq
This result is a corollary to the computation in \cite{Plek} of the
associated characteristic isometry $X\two$ satisfying the system of
identities (4.18), (4.19). The \vrep\ of this extension contains all
`diagonal' sectors of $\aa\two$ of the form $\pi_s \otimes \pi_{\bar
s}$ precisely once.

It is more convenient to deviate from the basis conventions in
\cite{Plek} and choose a $CPT$ conjugate pair of bases of isometric
intertwiners $T_e$ and $T_{\bar e} = j(T_e)$ on the two chiral \lc s
(cf.\ \cite{FLRR}). The anti-linear $CPT$ conjugation $j$ is an
appropriate Tomita-Takesaki modular conjugation \cite{FG,BGL}. It acts
like a reflection $x \leftrightarrow -x$ on the algebras of chiral
intervals, and relates conjugate sectors $\rho \leftrightarrow \brho =
j\comp\rho \comp j$. In such a basis, the isometry $X\two$ is simply
\be X\two = \Lambda^{-1/2} \sum_e \sqrt{\frac{d_td_s}{d_u}} \;
\tilde T_e \otimes \tilde T_{\bar e} \ee
where $\tilde T_e$ are local \itw s in $\aa\ch$ as in (4.17)
corresponding to the fusion channels $\rho_u \prec \rho_t\comp \rho_s$
as before, $\tilde T_{\bar e} = j(\tilde T_e)$ correspond to the $CPT$
conjugate channel $\brho_u \prec \brho_t\comp\brho_s$, and $d_s$ are the
statistical dimensions of $\rho_s$. The index equals $\Lambda = \sum_s
d_s^2$. The fusion channels contributing to the isometry $X\two$ for the
two-dimensional subtheory (4.3) are of the form $e \otimes \bar e$, and
the coefficients $\lam\two(e \otimes \bar e)$ are read off eq.\ (5.11).
The fact that the corresponding two-dimensional fields
\[ \Phi_s = \sum_{e \otimes \bar e} \sqrt\frac{d_td_s}{d_u} \;
  F(e \otimes \bar e, \eins\otimes\eins) \equiv \sum_e
 \sqrt\frac{d_td_s}{d_u} \; F(e,\eins) \otimes F(\bar e,\eins) \]
contracted from chiral exchange fields of fixed charge $[s]$, $[\bar s]$
are indeed local fields acting on the Hilbert space $\hh\two =
\bigoplus_t \hh_t \otimes \hh_{\bar t}$, was established in \cite{Spt}.
Although the diagonal sectors are not closed under composition whenever
there are non-simple fusion rules among the chiral sectors $\pi_s$, the
operator product of the diagonal fields $\Phi_s$ contains only other
diagonal fields due to cancellations among the fusion coefficients. This
is another instance of truncated fusion rules.

{}From
\[ \lam\two(e\otimes\bar e) = \sqrt\frac{d_s}{\Lambda}
           \sqrt\frac{d_t}{d_u}\]
it is obvious that the amplitudes for the 2D partial waves contributing
to a given $n$-point function of integer isospin fields $\LO, \Phi_n
\cdots \Phi_1 \RO = \sum_\xi N\two_{\xi\otimes\bar\xi} \;
\ff_\xi\cdot\bar\ff_{\bar \xi}$ are all equal:
\be N\two_{\xi\otimes\bar\xi} = \prod_i \sqrt{d_{s_i}/\Lambda}
                 \propto 1. \ee
Given the diagonal standard extension, we can predict characteristic
invariants for every other extension which can be read off the
respective $n$-point functions, independent of all normalizations of
partial waves and \cfb s, by taking double ratios of
amplitudes (5.5) and (5.12)
\[ \frac{(N_\xi/N_\eta)(N_{\bar\xi}/N_{\bar\eta})}
{N\two_{\xi\otimes\bar\xi}/N\two_{\eta \otimes \bar\eta}} = \prod_i
\frac{\lam(e_i)\bar\lam(\bar e_i)}{\lam(f_i)\bar\lam(\bar f_i)}
= \prod_i \frac{\vert\lam(e_i)\vert^2}{\vert\lam(f_i)\vert^2} . \]
Here we have used the fact that the coefficients of $X$ and $j(X)$
in $CPT$ conjugate bases are complex conjugates, $\bar\lam(\bar e) =
\ovl{\lam(e)}$. E.g., for the 4-point function of the isospin 3 field
in the $E_6$ model (4.2), we get
\be \frac{(N_3/N_0)^2}{N\two_3/N\two_0} = 2 \ee
in agreement with the result obtained previously (eq.\ (3.8) and
\cite{FGPFKS}) by the analysis of locality in terms of explicit \cfb\
functions given as solutions to KZ differential equations.

We emphasize that this method works for every `non-diagonal' extension
of a given chiral theory without controlling the actual pointlike limits
$F(e,a) \arr \vphi_e(x)$, since there is always the `diagonal' one to
compare with. Moreover, it immediately applies to mixed and higher
$n$-point functions.

We conclude this section with another instructive (albeit almost
trivial) example giving rise to anyonic field extensions. We consider a
local theory $\aa$ with $N$ simple \sps\ sectors $\rho_s$ with $\ZZ_N$
fusion rules $[s] [t] = [s+t \; ({\rm mod} N)]$. For simplicity, assume
that the \amor s $\rho_s$ can be chosen to satisfy $\rho_s\rho_t =
\rho_{s+t}$ (understood mod $N$), by which all \itw s $T_e$ of the
general analysis are trivial $=\eins$. This choice is always possible
for odd $N$, and for even $N$ provided the fractional spin of $\rho_s$
satisfies $N\Delta_s \in \ZZ$ (cf.\ \cite{Spt}). The sector structure
is that of the simple sectors in $su(N)$ \ca s. It also occurs in the
models constructed in \cite{BMT}, where, however, the violation of the
spin condition leads to a minor complication which we want to ignore
here. The case $N=2$ includes the $D_n$ series of chiral $su(2)$ \ca\
extensions.

We choose a complete system of orthonormal isometries $W_s$ and
construct the reducible \emor\ $\rho(a) := \sum_s W_s \rho_s(a) W_s^*$.
Then the triple $(\rho,W,X)$ where $W=W_0$ and
\be X := N^{-1/2} \sum_{st} \rho(W_s) W_t W_{s+t}^* \ee
(with trivial \CGC s for an abelian group) solves the system (4.18).
The charged fields $\psi_s$ are obtained (up to a
normalization factor $N^{1/2}$) as the unitary shift operators
$|t;\Psi\rangle \mapsto |t+s;\Psi\rangle$ on $\bigoplus_t \hh_t$.
They satisfy $\psi_s\psi_t = \psi_{s+t}$ and implement the \emor s
$\rho_s$ (in the \rep\ $\pi^0 = \bigoplus \pi_s$)
\be \rho_s(a) = \psi_s a \psi_s^* \qquad (a \in \aa).\ee
The gauge group $\ZZ_N$ acts by $\g_n(\psi_s) = e^{2\pi i ns/N}
\psi_s$ with average $\mu(\psi_s) = \d_{s0} \eins$. Putting
\be V := N^{-1/2}\sum_s W_s \psi_s, \ee
and defining $\g$ by (4.6) with index $\lam = \vert\ZZ_N\vert = N$,
then $\g(V) = X$ and the triple $(\g,V,W)$ satisfies the identities
(4.4). Adjoining the charged fields $\psi_s$ to the local algebras, we
obtain an anyonic field extension $\bb$ by the simple sectors of $\aa$.
\sectreset{Concluding remarks}
The old hope that the `germ of the observable algebra' generated by the
internal symmetry currents and the \emt\ completely determines a local
\qft\ turns out to require some qualifications. Two-dimensional
conformal \ca\ models tell us that depending on the value of the level
$k$ (which characterizes both the algebra $\aa_k$ and the vacuum state
of the theory), there may be several -- one, two, or three for
$\aa_k(su(2))$ -- local conformal field theories corresponding to the
same \vrep\ of $\aa_k$.

The different theories are distinguished by different maximal local
chiral extensions $\bb_k$ and by different braid invariant quadratic
forms $M$. The primary local chiral fields which extend $\aa_k$ obey
fusion rules which are majorized by the intrinsic DHR fusion rules of
\sps\ sectors. Both the invariant ratios of structure constants which
are characteristic quantities for local field extensions, and the
truncated fusion rules are understood and computed in conventional field
theoretical terms and in terms of the theory of subfactors applied to a
single local subfactor $\aa(\ii) \subset \bb(\ii)$.

Our field theoretical computation uses a closed expression for the
$s$-channel fusion matrix (that is already implicit in \cite{STHI})
which has the virtue of displaying their invariance under Galois \amor s
(the individual structure constants as well as the matrix elements of
the monodromy \rep\ of the mapping class group belonging to the same
algebraic number field). The relevance of such arithmetic properties has
been recently exhibited in a study of the Schwarz problem (`When is the
\rep\ of the braid group a finite matrix group?') for the KZ equation
\cite{ST}.

On the other hand, the application of the theory of finite index
subfactors to local field extensions gives a natural interpretation of
the field theoretical structures in terms of a generalized `harmonic
analysis'. The `irreducible tensor operators' of this analysis are
the quantum field theoretical charged intertwiners. This approach is
very close to the spirit of Ocneanu who first considered subfactors as
`generalized groups', but gives more evidence to this view than the
combinatorial description in terms of bi-partite graphs and connections
\cite{Oc}. Part of Ocneanu's induction-restriction graph is reflected
in the `truncated fusion rules' which in turn derive from harmonic
analysis in the form of operator product expansions for charged fields.
Through Longo's theorem relating the truncation to the depth of the
inclusion \cite{Hopf}, it is nicely exhibited that the generalized
symmetry associated with conformal embeddings is not given by a Hopf
$C^*$ algebra in general. Longo's characterization of a subfactor in
terms of a triple $(\rho,W,X)$ gives rise to a notion of generalized
\CGC s which does not refer to any assumed linear transformation law of
the irreducible tensor operators. We note that the interpretation of
these structures as a generalized symmetry is not imposed but emerges
naturally from the theory of subfactors.

When one compares our two different approaches, one can also observe
some unbalance. E.g., the role of the Galois \amor s is not yet
understood in terms of the subfactor approach. In particular, the Galois
group acting on the structure constants does not map a unitary theory
into another unitary theory, nor are there any `Galois relatives' of a
subfactor. Indeed, the characteristic ratios of structure constants like
(3.8), (3.9), (3.15) resp.\ (5.13) turn out to be rational numbers and
are, therefore, Galois invariants.

The characterization of a local extension in terms of a triple
$(\rho,W,X)$ as in Prop.\ 4.2 logically proceeds in two steps: first,
one has to solve the system (4.18) which, among other things, controls
the consistent truncated operator product expansions. This already
yields field extensions which, however, may be non-local. E.g., a
fermionic field theory as an extension of its even (bosonic) subtheory
arises in this way. The locality condition (4.19) is only imposed in a
second step.
On the other hand, in the \cfb\ approach the locality condition seems
to be the only vital step. In fact, we consider the analogue of the
first step to be hidden in the KZ equation, whose solutions
automatically give rise to a consistent fusion.
\vspace{9mm} \par
\addtolength{\baselineskip}{-\deltaabs}
{\bf Acknowledgements:} The discussions that gave rise to the present
paper started when all three authors were visiting the Erwin
Schr\"odinger International Institute for Mathematical Physics in Vienna
in early 1993. Most of the work was done during a second visit at the
ESI by Ya.S.\ and I.T.\ in the spring of 1994. We all benefited from the
hospitality of the collegues in Vienna and of the creative atmosphere at
the Schr\"odinger Institute. I.T.\ thanks J.-B.\ Zuber for a
stimulating discussion on the special properties of ratios of structure
constants, and for acquainting him with the work in progress of V.\
Petkova and himself \cite{PZ} that relates these constants to the
eigenvectors of associated Cartan matrices. Ya.S.\ and I.T.\
acknowledge partial support by the Bulgarian Foundation for Scientific
Research under contract F-11.
\def\AP#1{Ann.\ Phys.\ (N.Y.) {\bf #1}}
\def\CMP#1{Com\-mun.\ Math.\ Phys.\ {\bf #1}}
\def\IM#1{Invent.\ Math.\ {\bf #1}}
\def\JFA#1{J.\ Funct.\ Anal.\ {\bf #1}}
\def\JMP#1{Journ.\ Math.\ Phys.\ {\bf #1}}
\def\LMP#1{Lett.\ Math.\ Phys.\ {\bf #1}}
\def\LNP#1{Lect.\ Notes Phys.\ {\bf #1}}
\def\NP#1{Nucl.\ Phys.\ {\bf #1}}
\def\PL#1{Phys.\ Lett.\ {\bf #1}}
\def\PR#1{Phys.\ Rev.\ {\bf #1}}
\def\PRL#1{Phys.\ Rev.\ Lett.\ {\bf #1}}
\def\RMP#1{Rev.\ Math.\ Phys.\ {\bf #1}}
\def\etal{{\it et al.}}
\def\KHR{K.-H.\ Rehren} \def\BS{B.\ Schroer} \def\KF{K.\ Fredenhagen}
\def\JF{J.\ Fr\"ohlich} \def\DB{D.\ Buchholz} \def\RH{R.\ Haag}
\def\SD{S.\ Doplicher} \def\JR{J.E.\ Roberts} \def\RL{R.\ Longo}
\def\WS{World Scientific} \def\AMS{Am.\ Math.\ Soc.}
\small \addtolength{\baselineskip}{-\deltaref}

\end{document}